# Phase diagram of superconducting vortex ratchet motion in a superlattice with noncentrosymmetry

*Shengyao Li,[1] Lijuan Zhang,[2] Ke Huang,[1] Chen Ye,[1] Tingjing Xing,[3] Liu Yang,[4] Zherui Yang,[1]Qiang Zhu,[5] Bo Sun,[2,6] Xueyan Wang,[1*] X. Renshaw Wang (王骁)[1,7*]*

[1]Division of Physics and Applied Physics, School of Physical and Mathematical Sciences, Nanyang Technological University, 21 Nanyang Link, Singapore 637371, Singapore

[2]Tsinghua-Berkeley Shenzhen Institute and Shenzhen Geim Graphene Center, Tsinghua University, Shenzhen, Guangdong 518055, China.

[3]Murray Edwards College, University of Cambridge, Huntingdon Rd, Cambridge CB3 0DF, United Kingdom

[4]Raffles Institution, 1 Raffles Institution Lane, Singapore 575954, Singapore

[5]Institute of Materials Research and Engineering (IMRE), A*STAR, 2 Fusionopolis Way, 138634 Singapore

[6]Institute of Materials Research, Tsinghua Shenzhen International Graduate School, Guangdong Provincial Key Laboratory of Thermal Management Engineering and Materials, Shenzhen, Guangdong 518055, China

[7]School of Electrical and Electronic Engineering, Nanyang Technological University, 50 Nanyang Ave, 639798, Singapore

E-mail: renshaw@ntu.edu.sg
xueyan.wang@ntu.edu.sg

## Abstract

Ratchet motion of superconducting vortices, which is a directional flow of vortices in superconductors, is highly useful for exploring quantum phenomena and developing superconducting devices, such as superconducting diode and microwave antenna. However, because of the challenges in the quantitative characterization of the dynamic motion of vortices, a phase diagram of the vortex ratchet motion is still missing, especially in the superconductors with low dimensional structures. Here we establish a quantitative phase diagram of the vortex ratchet motion in a highly anisotropic superlattice superconductor, $(SnS)_{1.17}NbS_2$, using nonreciprocal magnetotransport. The $(SnS)_{1.17}NbS_2$, which possesses a layered atomic structure and noncentrosymmetry, exhibits nonreciprocal magnetotransport in a magnetic field perpendicular and parallel to the plane, which is considered a manifest of ratchet motion of superconducting vortices. We demonstrated that the ratchet motion is responsive to current excitation, magnetic field and thermal perturbation. Furthermore, we extrapolated a giant nonreciprocal coefficient ($\gamma$), which quantitatively describes the magnitude of the vortex ratchet motion, and eventually established phase diagrams of the ratchet motion of the vortices with a quantitative description. Last, we propose that the ratchet motion originates from the coexistence of pancake vortices (PVs) and Josephson vortices (JVs). The phase diagrams are desirable for controlling the vortex motion in superlattice superconductors and developing next-generation energy-efficient superconducting devices.

**Introduction**

Ratchet motion is a type of intriguing and useful motion allowed in one direction but prevented in the opposite direction. In the research field of condensed matter physics, the superconducting (SC) vortices could exhibit the ratchet motion when they are driven by the Lorentz force in superconductors with noncentrosymmetric structures [1–3]. Appealingly, the ratchet motion of vortices shows properties [4,5] important for designing future superconductor-based devices, such as superconducting diode and microwave antenna [6–13].

In a superconducting system, two kinds of vortices, namely disc-like pancake vortices (PVs) and elliptical Josephson vortices (JVs), could be generated. In particular, the elliptical JV solely occurs in layered superconductors with alternating superconducting and non-superconducting layers. To be specific, PVs penetrate the superconducting layers, while JVs are sandwiched between the superconducting layers [2,14]. Additionally, the ratchet motion of PVs and JVs are also dynamically responsive to vortex density, driving force and pinning potential [15], potentially contributing to not only the design of superconducting devices but also the discovery of novel physical phenomena [16,17]. Therefore, a detailed quantitative study of the ratchet motion of superconducting vortices is of great interest.

A fascinating platform for capturing the vortex dynamics of the vortex rachet motions is the noncentrosymmetric superlattice superconductors, such as $(SnS)_{1.17}NbS_2$ [18–21]. Because of the noncentrosymmetry of the atomic layers, vortices are expected to exhibit ratchet motion in the $(SnS)_{1.17}NbS_2$ [1,22,23]. Recently, the advancement of nonreciprocal magnetotransport demonstrated that the vortex ratchet motion could be quantitatively revealed by measuring the second-harmonic magnetoresistance ($R^{2\omega}$) of the superconducting system. Thus far, the nonreciprocal magnetotransport has been observed in noncentrosymmetric superconductors, such as interfacial/polar superconductors [22,24], transition metal dichalcogenide superconductors with a strong polarity [3,25], magnet-superconductor heterostructures [26] and artificially-asymmetric superconducting array [9,27]. Since the dynamic characterization of JV's ratchet motion is elusive and $R^{2\omega}$ is a good indicator to differentiate the ratchet motion of various vortices, we employ the nonreciprocal magnetotransport to construct phase diagrams of vortex's ratchet motion.

Here we quantitatively studied the vortex ratchet motion in $(SnS)_{1.17}NbS_2$. We observed a nonreciprocal magnetotransport during the superconducting transition in the magnetic field perpendicular and parallel to the sample plane. Each peak in the magnetic field-dependent nonreciprocal $R^{2\omega}$ represents a particular vortex ratchet motion with a distinct nonreciprocal coefficient ($\gamma$). The vortex ratchet motion can be tuned by current excitation, magnetic field and thermal perturbation. Considering the alternate stack of superconducting and non-superconducting layers, we attribute this nonreciprocal magnetotransport in magnetic fields of different directions to the ratchet motion of vortices. Moreover, the ratchet motion in a perpendicular magnetic field exhibits a giant $\gamma$, which is several orders higher than most of the records.

**Results & discussion**

Figure 1(a) shows the structural illustration of $(SnS)_{1.17}NbS_2$, where red, orange and green spheres represent Nb, Sn and S, respectively. The blue dashed-line box denotes a unit cell, where the structural motif of SnS and $NbS_2$ are 4 and 2, respectively [21]. The sublayers share similar *b*- and *c*-lattice constants, while the ratio of the *a*-lattice constants between $NbS_2$ and SnS sublayers is ~0.585 [20]. The detailed structural illustration and calculation can be found in Fig. S1. High-resolution transmission electron microscopy (HRTEM) confirms the layered structure and distinguishes the two sublayers by the different contrasts, with the SnS sublayers appearing brighter and the $NbS_2$ layers darker in Fig. 1(b). No obvious turbostratic disorder was observed from the HRTEM image, indicating high structural quality. Figure 1(c) shows the X-ray diffraction (XRD) pattern of an $(SnS)_{1.17}NbS_2$. The Bragg maxima are narrow and sharp, and they are indexed to the (001) family of planes, indicating a good crystallographic alignment of the layers. The $(SnS)_{1.17}NbS_2$ flakes were exfoliated and patterned into a Hall bar device (inset of Fig. 1(d)). The thickness of the device is 93 nm (Fig. S2), which can be regarded as bulk. We first characterized the superconductivity of our $(SnS)_{1.17}NbS_2$. The device shows a metallic behaviour upon cooling (Fig. 1(d)), and it shows a superconducting transition at $T_c$ =2.75 K, where $T_c$ is defined as the intersection of the two red dashed lines in Fig. 1(e). The temperature-dependent normalized resistance $R/R$(3 K) under $c$-axis magnetic fields shows a residual resistance under small magnetic fields. The broadening of superconducting transition in an external magnetic field is a signature of the vortex effect and is frequently seen in highly anisotropic layered superconductors [28,29]. To demonstrate the anisotropy of $(SnS)_{1.17}NbS_2$, the angular dependence of the upper critical field, $\mu_0 H_{c2}$, is displayed in Fig. 1(f). $\mu_0 H_{c2}$ increases abruptly as the magnetic field rotates from out-of-plane (OOP) to in-plane (IP) direction and shows a sharp cusp for IP magnetic field. The angular-dependent $\mu_0 H_{c2}$ can be well fitted by the 2D Tinkham model. This is in contrast to the anisotropic 3D character of 2$H$-$NbS_2$ [29], and con``firms the two-dimensional (2D) superconductivity in $(SnS)_{1.17}NbS_2$.

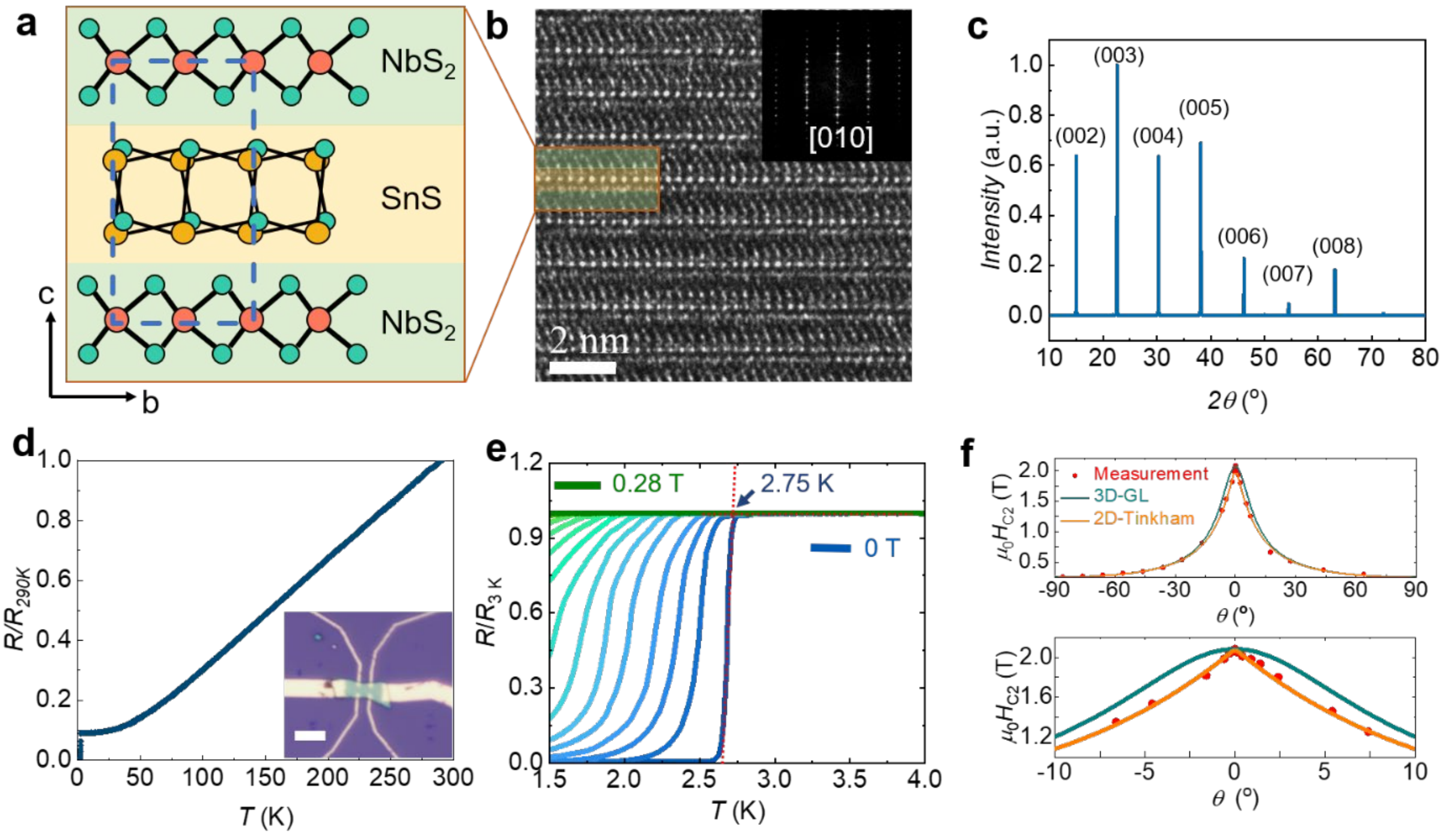

FIG. 1. Crystal structure and superconductivity of $(SnS)_{1.17}NbS_2$. (a) Crystal structure of $(SnS)_{1.17}NbS_2$, where red, orange, and green balls stand for Nb, Sn and S. The dashed-line box denotes a unit cell. (b) Cross-sectional HRTEM image along [010] direction. (c) XRD pattern of a single crystal of $(SnS)_{1.17}NbS_2$. (d) Temperature dependence of the normalized resistance $R/R_{290\,K}$ from room temperature to 1.5 K. Inset is an optical image of the Hall bar device. Scale bar: 6 μm. (e) Normalized temperature-dependent resistance $R/R$(3 K) in a $c$-axis magnetic field with a step of 0.02 T. (f) Angular dependence of upper critical field, $\mu_0 H_{c2}$ ( $\theta$ ), at 1.6 K. $\mu_0 H_{c2}$ is defined as the field where the 95% of normal resistance is reached.

The fundamental structure of $(SnS)_{1.17}NbS_2$ is the monolayer of 1*H*-$NbS_2$ sandwiched by SnS layers, as shown in Fig. 2(a). The inversion symmetry is broken by the trigonal prismatic coordination of S atoms around the Nb atom in 1*H*-$NbS_2$, thus the whole system is an ideal system to study the ratchet motion and the dynamics of vortices. In order to comparatively study magnetotransport characteristics in different bulk superlattices induced by broken inversion symmetry, we also fabricated devices of $(SnS)_{1.17}(NbS_2)_2$. Figure 2(b) displays the crystal structure of $(SnS)_{1.17}(NbS_2)_2$. Compared with $(SnS)_{1.17}NbS_2$, it shares a similar stacking order, with the bilayer of 2*H*-$NbS_2$ being sandwiched by SnS layers. However, the bilayer 2*H*-$NbS_2$ exhibits the inversion symmetry, as shown in the dashed line box in Fig. 2(b). When a superconductor is exposed to the magnetic field, partial magnetic flux can penetrate the superconductors, forming a quantized vortex circulated by current loops. These vortices can be driven into flow motion by the Lorentz force [4,30]. Especially, an IP magnetic field may penetrate the layered superconductor in the form of JV columns parallel to the SnS sublayers. And the OOP field component penetrates in the form of disk-like PVs perpendicular to the superconducting sublayers, as shown in Fig. 2(c) [31–34]. Owing to the similar stacking order, both PVs and JVs in layered superconductors flow along the *ab* plane by tuning the direction of the magnetic fiel [35,36]. Hence, the ratchet motion of vortices is possible in $(SnS)_{1.17}NbS_2$ due to the broken inversion symmetry.

Recently, the measurements of nonreciprocal magnetoresistance have been developed to investigate vortex dynamics [3,22,23]. The ratchet motion of vortices produces a current- and magnetic field direction-dependent nonreciprocal $R^{2\omega}$. Therefore, this $R^{2\omega}$ confidently reveals the characteristics of vortex ratchet motion (see Part 2 in Supplemental Materials) [25,37]. Figures 2(d) and 2(e) display $R^{\omega}$-$B$ curves of $(SnS)_{1.17}NbS_2$ and $(SnS)_{1.17}(NbS_2)_2$, respectively, with the magnetic field varying from IP ($\theta$ =0°) to OOP ($\theta$ =90°) directions at 1.6 K. Both $(SnS)_{1.17}NbS_2$ and $(SnS)_{1.17}(NbS_2)_2$ are highly anisotropic with $\mu_0H_{c2}$ (IP) approximately ten times higher than $\mu_0H_{c2}$(OOP). Interestingly, during the superconducting transition, the $R^{\omega}$-$B$ curves of $(SnS)_{1.17}(NbS_2)_2$ have the same curvature in the magnetic field of different directions (Fig. 2(e)), while the curvature of $R^{\omega}$-$B$ curves of $(SnS)_{1.17}NbS_2$ varies with the magnetic field directions (Fig. 2(d)). This could be a consequence of the directional motion of vortices in $(SnS)_{1.17}NbS_2$. Figure 2(f) shows the $R^{2\omega}$ of $(SnS)_{1.17}NbS_2$ measured simultaneously with $R^{\omega}$. With the application of an OOP magnetic field, $R^{2\omega}$ is negligibly small at the zero-resistance state and suddenly rises with a steep peak during the superconducting transition. This is in agreement with the previous studies of the nonreciprocal transport in noncentrosymmetric superconductors [3,11,23–25,38]. With the magnetic field rotating away from the OOP direction, the $R^{2\omega}$ gradually decreases. Interestingly, another $R^{2\omega}$ signal with an opposite sign emerges and reaches its maximum in an IP magnetic field. In contrast, Figure 2(g) shows that $(SnS)_{1.17}(NbS_2)_2$ does not show $R^{2\omega}$ responses in magnetic fields of any direction in a device with similar geometric structure and thickness (see Fig. S10). Considering the similar structure of the two crystals, *i.e.* $(SnS)_{1.17}NbS_2$ and $(SnS)_{1.17}(NbS_2)_2$, we speculate that the different $R^{2\omega}$ responses in magnetic fields of various directions may originate from the ratchet motion of different vortices. Namely, PV contributes to the $R^{2\omega}$ response in the OOP and tilted magnetic field, JV contributes to the $R^{2\omega}$ response in an IP magnetic field, based on the following four reasons.

First, the $R^{2\omega}$ under an IP magnetic field does not originate from the interface-induced Rashba effect [13]. Our $(SnS)_{1.17}NbS_2$ and $(SnS)_{1.17}(NbS_2)_2$ have identical SnS-$NbS_2$ interfaces, thus could have similar interfacial polarity. However, $(SnS)_{1.17}(NbS_2)_2$ does not show any $R^{2\omega}$ response in magnetic fields of any direction, excluding the possibility of the SnS-$NbS_2$ interfaces that induced $R^{2\omega}$ responses. Second, the $R^{2\omega}$ under an IP magnetic field is not a contribution of tilted PV chains [35,39,40]. The unique $R^{2\omega}$ $(\theta = 0^{\circ})$ only exists for a small range of tilted magnetic fields $(\theta = \pm 5^{\circ})$, as shown in Fig. 2(f). This is distinct from the tilted vortices model, which lasts for a wide range of tilted magnetic fields. Third, the $R^{2\omega}$ under an IP magnetic field is not from PV bending [41]. Vortex usually bends at the sample defects and edges under the Lorentz force. But PV is as flat as 1 nm, which does not correspond to a long vortex bend model. In addition, the appearance of $R^{2\omega}$ responses is irrelevant to the strength of Lorentz force (Fig. S9), which is different from the strongly Lorentz force-correlated vortex bending effect. Last, the $R^{2\omega}$ is not a contribution from defect or geometric potentil [9]. The result is repeatable for samples of different geometric dimensions, but the defects and geometry potential are randomly distributed. Therefore, the $R^{2\omega}$ responses in $(SnS)_{1.17}NbS_2$ are likely due to the ratchet motion of PVs and JVs in the magnetic field of different directions.

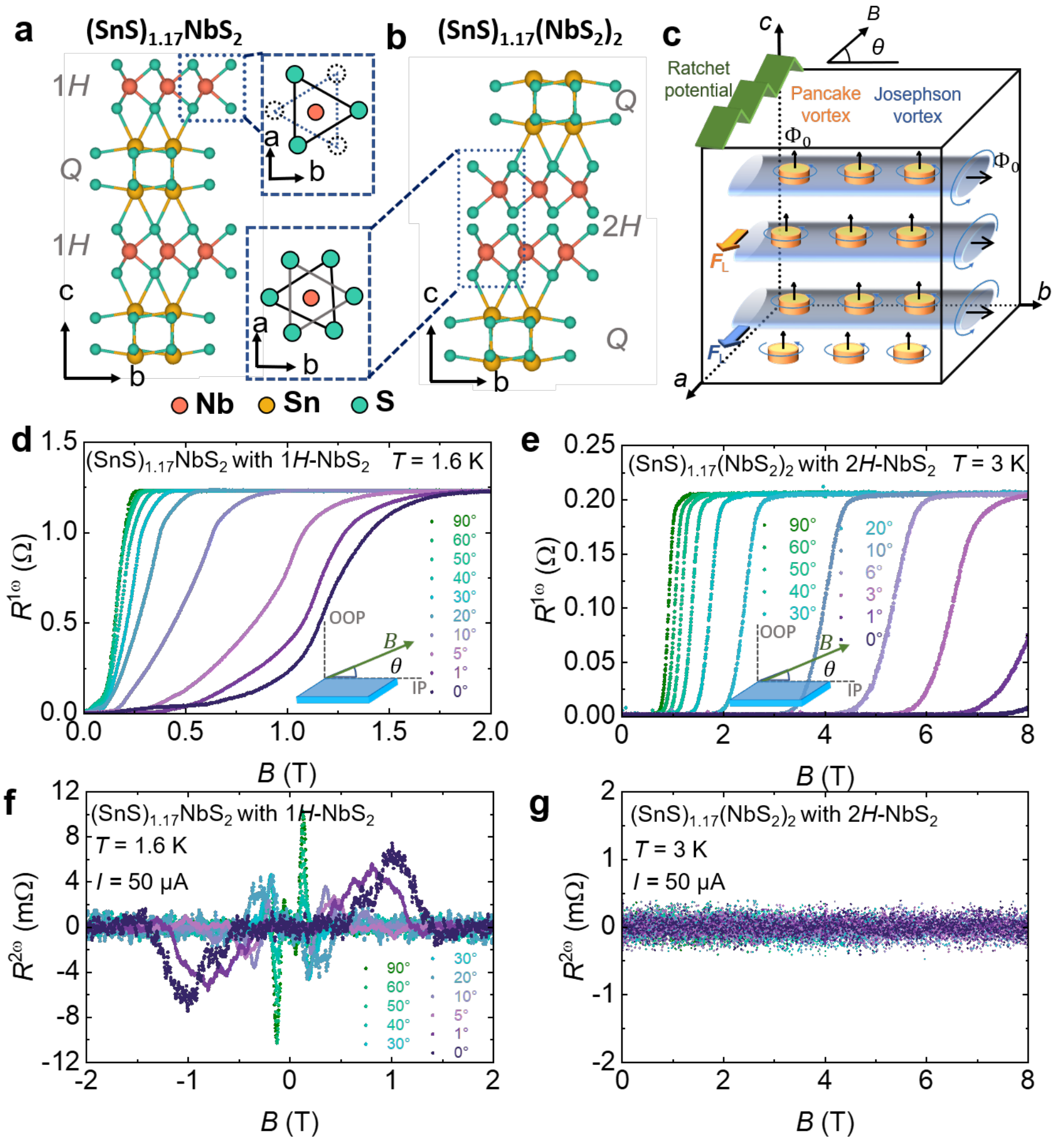

FIG. 2: Magnetotransport of 1*H*-$NbS_2$- and 2*H*-$NbS_2$-based superlattice. (a,b) Crystal structures of (a) $(SnS)_{1.17}NbS_2$ with 1*H*-$NbS_2$ interlayer and (b) $(SnS)_{1.17}(NbS_2)_2$ with 2*H*-$NbS_2$ interlayer. (c) Schematic illustration of coexistence of PVs and JVs in layered superconductors in the applied magnetic field. (d) First harmonic magnetoresistance, $R^{\omega}$, of $(SnS)_{1.17}NbS_2$ as a function of magnetic fields at $T$ = 1.6 K. $\theta$ is defined as the angle between the magnetic field and the sample plane. (e) $R^{\omega}$ of $(SnS)_{1.17}(NbS_2)_2$ as a function of magnetic fields at $T$ = 3 K. (f) $R^{2\omega}$ of $(SnS)_{1.17}NbS_2$. (g) $R^{2\omega}$ of $(SnS)_{1.17}(NbS_2)_2$.

Because the Lorentz force drives the plastic flow of vortices, current- and temperature-dependent nonreciprocal magnetotransport was carried out to explore the ratchet motion under different driving forces and pinning potential, respectively [42,43]. Figure 3 shows the current amplitude-dependent vortex ratchet motion in OOP and IP magnetic fields. We build the phase diagram of PV ratchet motion (OOP magnetic field) against current amplitude and magnetic field strength accordingly in Fig. 3(a). The $R^{\omega}$ can be found in Fig. S3. Upon the application of a magnetic field, the sample first stays in the Meissner state and then the vortex solid (V solid) state without the $R^{2\omega}$ response. With the increase of the magnetic field, the flow of vortices shows a ratchet motion ($R^{2\omega}$ responses) in the resistive state. Last, the sample goes to the normal state when superconductivity is destroyed. Here, we observed a sign reversal of $R^{2\omega}$ with the increase of the magnetic field. This phenomenon has been attributed to the vortex proliferation-induced drift direction chang [27]. Thus, we define the two regions as low-density of PVs (LD-PV) and high-density of PVs (HD-PV) accordingly. LD-PV gradually emerges as the current amplitude is increased above 50 μA and can be transformed to HD-PV by increasing the magnetic field strength. This conversion indicates that the current amplitude and magnetic field strength can modify the nonreciprocity of the ratchet motion of PVs. In detail, a higher current is needed for the ratchet motion of LD-PV and a lower current for HD-PV. The absolute maximum values of $R^{2\omega}$ response in HD-PV, $|R^{2\omega}_{\mathrm{MAX}}|$, are extracted as a function of current amplitude (see Fig. 3(b)). The vortex ratchet motion manifests the dome-like shape due to the competitive effect between the pinning strength and the driving force [27,44]. Increasing current below 50 μA benefits the vortex ratchet motion with an increased driving force, but excessive current above 50 μA weakens the pinning site and decreases the ratchet effect. In contrast, the ratchet motion of JVs (IP magnetic field) remains the single state with a constant nonreciprocity, regardless of the variations in current amplitudes (Fig. 3(c)). The $|R^{2\omega}_{\mathrm{MAX}}|$ as a function of current amplitude in an IP field also exhibits a current-related dome-like shape (see Fig. 3(d)), suggesting that JVs share a similar pinning effect as PVs.

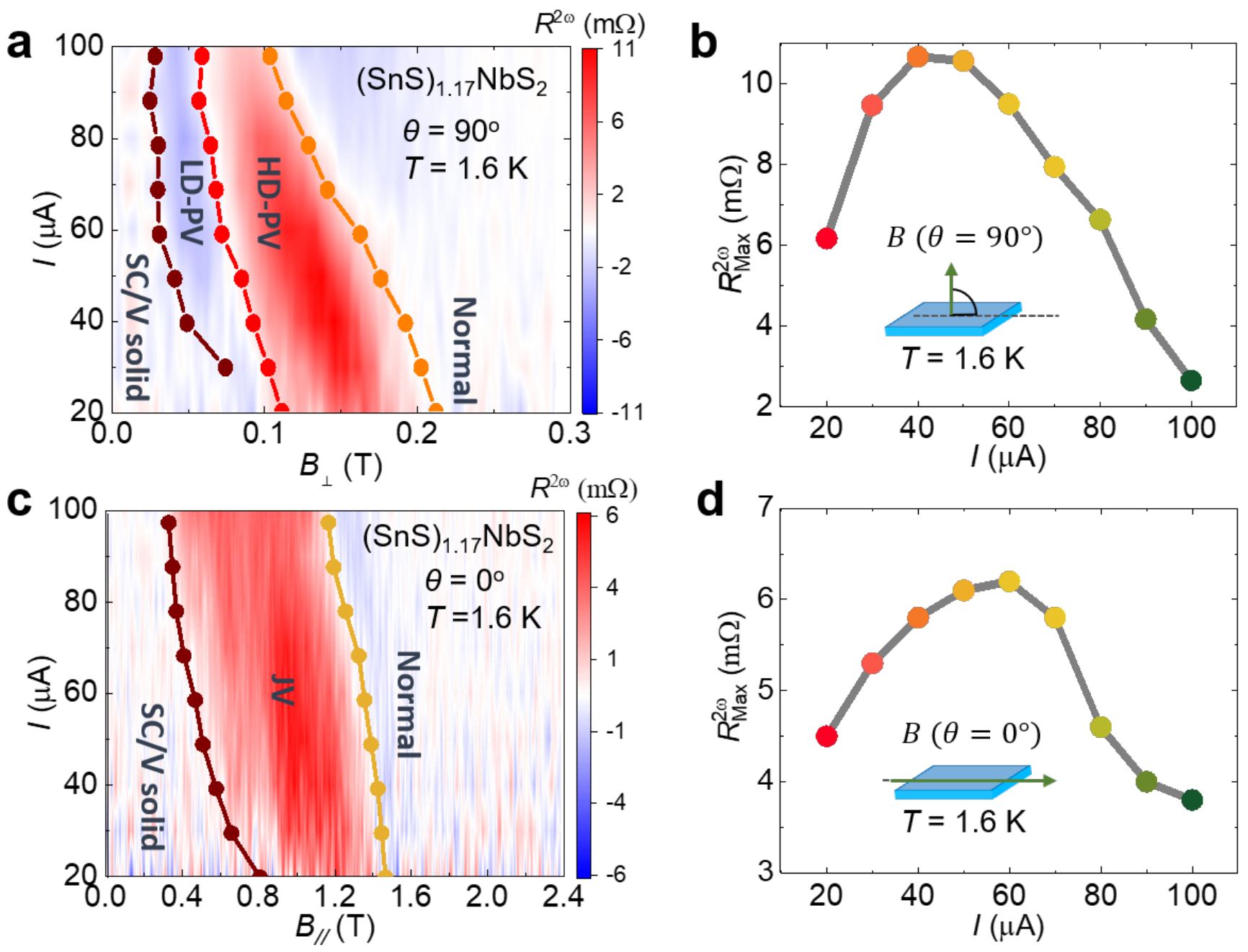

FIG. 3: Quantitative current-dependent phase diagrams of vortex ratchet motion in OOP and IP magnetic fields. (a) Phase diagram of PV ratchet motion in the $I$-$B_{\perp}$ plot at 1.6 K. The dots represent the emergence and the disappearance of $R^{2\omega}$ responses, marking the transition between neighbouring phases. (b) The extracted absolute maximum value, $|R^{2\omega}_{\mathrm{MAX}}|$, as a function of current in an OOP magnetic field. (c) Phase diagram of JV ratchet motion in the $I$-$B_{//}$ plot at 1.6 K. (d) The extracted $|R^{2\omega}_{\mathrm{MAX}}|$ as a function of current amplitude in an IP magnetic field.

Thermal perturbation dramatically influences the motion behaviour of vortices by weakening the pinning strength at the defects, therefore, promoting the plastic flow of vortices [45–47]. In addition, with the temperature approaching $T_c$, we observed an abrupt emergence of a new $R^{2\omega}$ response. This phenomenon was attributed to the thermally-assisted flux flow (TAFF) behaviour [11,48–50]. Therefore, we could further obtain a phase diagram of vortex ratchet motion in a *T*-*B* plot. Figure 4(a) shows the PV ratchet motion in the $T$-$B_\perp$ plot. At a low temperature and a low field, vortices are pinned at asymmetric pinning potentials without $R^{2\omega}$ response. The increases in magnetic fields and temperatures promote the flowing of vortices by overcoming the pinning potential, with a transition from vortex solid state to vortex liquid state. At 1.6 K and under a current of 50 μA, 0.46 T is needed for the start of the ratchet motion of JVs in an IP magnetic field, and 0.04 T for the PVs in an OOP magnetic field. The nucleation of PVs causes the transition from LD-PV to HD-PV [27], and the melting behaviour of the pinned vortices in the TAFF state is responsible for the transition from HD-PV to TAFF approximately at $T_c$ [3,45]. In order to quantitatively capture the ratchet motion, the nonreciprocal coefficient, $\gamma$, is subsequently calculated as $\gamma = \frac{2|R_{MAX}^{2\omega}|}{BR^{\omega}I_0}$ at each temperature, where $|R_{\mathrm{MAX}}^{2\omega}|$ is the maximum value of $R^{2\omega}$ with the corresponding values of magnetic field $B$ and $R^{\omega}$. We define the nonreciprocal coefficient in the HD-PV and TAFF states as $\gamma_1 = \frac{2|R_{MAX1}^{2\omega}|}{BR^{\omega}I_0}$ and $\gamma_2 = \frac{2|R_{MAX2}^{2\omega}|}{BR^{\omega}I_0}$, respectively. Figure 4(b) shows the temperature-dependent $\gamma_1$ and $\gamma_2$. $\gamma_1$ first increases with the rising temperature, then stabilizes with a maximum value of ~$2.00\times10^4$ $A^{-1}T^{-1}$, and finally quenches above 2.4 K. $\gamma_2$ emerges at 2.2 K, then suddenly rises to the maximum of $4.04\times10^4$ $A^{-1}T^{-1}$ at 2.6 K, and finally quenches above $T_c$. In comparison with the previously reported noncentrosymmetric superconducting systems, such as $Bi_2Te_3$/FeTe interface ($\gamma$ ~$10^{-3}$ $A^{-1}T^{-1}$) [24], ionic liquid gated $MoS_2$ thin flake ($\gamma$ ~$10^3$ $A^{-1}T^{-1}$) [3], atomically thin $NbSe_2$ film ($\gamma$ ~$10^2$ $A^{-1}T^{-1}$) [11], artificial superlattice $[Nb/V/Ta]_n$ ($\gamma$ ~$10^2$ $A^{-1}T^{-1}$) [13]. Our giant $\gamma$ is several orders higher than most of the previous records, and it proves the great potential of the bulk superlattice superconductors to achieve the giant nonreciprocal electrical transport. On the contrary, the temperature-dependent $R^{2\omega}$-$B$ curves in an IP field exhibit a single pair of peaks regardless of the magnetic field strength and temperature variations. Figure 4(c) shows the phase diagram of JV ratchet motion in the $T$-$B_{//}$ plot, where the JV ratchet motion remains a constant nonreciprocity in the whole temperature range. The corresponding $\gamma$ at each temperature is calculated in Fig. 4(d), which is smaller than that in an OOP field. Therefore, we conclude that the JV ratchet motion with a consistent nonreciprocity is stable under the excitation of current, magnetic field and temperature perturbation.

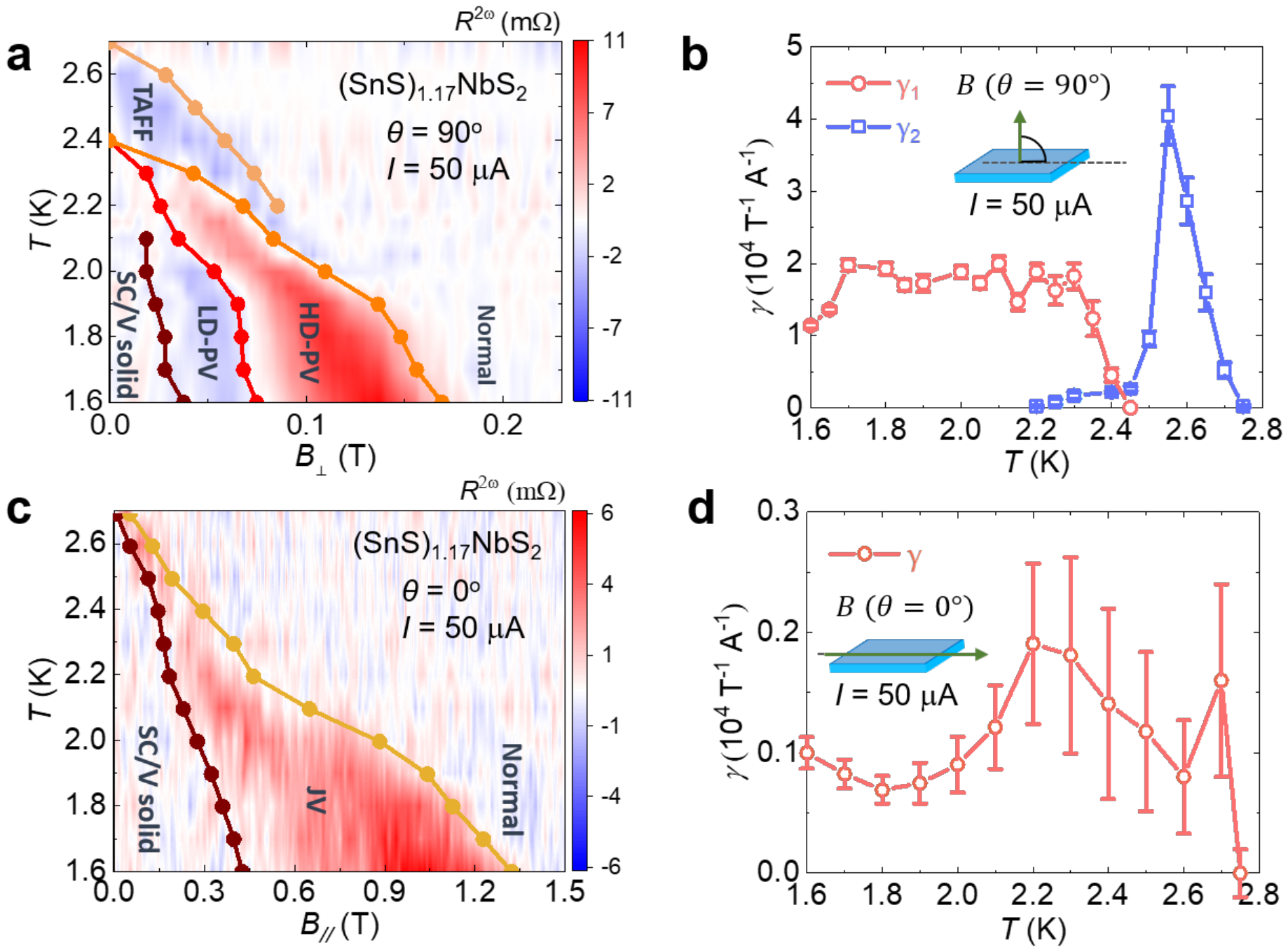

FIG. 4: Quantitative temperature-dependent phase diagrams of vortex ratchet motion in OOP and IP magnetic fields. (a) Phase diagram of PV ratchet motion in the $T$-$B_{\perp}$ plot. (b) Temperature-dependent $\gamma_1 = \frac{2|R^{2\omega}_{MAX1}|}{BR^{\omega}I_0}$ and $\gamma_2 = \frac{2|R^{2\omega}_{MAX2}|}{BR^{\omega}I_0}$ in an OOP magnetic field. (c) Phase diagram of JV ratchet motion in the $T$- $B_{//}$ plot. (d) Temperature-dependent $\gamma$ in an IP magnetic field.

## Conclusion

In summary, we have quantitatively studied the superconducting vortex ratchet motion in a noncentrosymmetric superlattice superconductor $(SnS)_{1.17}NbS_2$ as functions of the magnetic field, excitation current and thermal perturbation. Through the measurement of the nonreciprocal $R^{2\omega}$, each peak corresponds to a particular vortex ratchet motion with a unique $\gamma$. By comparing the lattice structure and transport characteristics of $(SnS)_{1.17}NbS_2$ and $(SnS)_{1.17}(NbS_2)_2$, we confirmed the nonreciprocal magnetoresistance in $(SnS)_{1.17}NbS_2$ is induced by the vortex motion in inversion broken 1$H$-$NbS_2$. We propose that, with the magnetic field turned from IP to OOP directions, multiple vortex arrangements of PVs and JVs are formed due to the layered structure with distinctive nonreciprocity. The PV ratchet motion can be separated into several phases by varying magnetic field, current and temperature, with a giant $\gamma$ during the superconducting transition. In contrast, the JV ratchet motion shows extra stability to external stimuli. The distinctive phase diagram of PV and JV ratchet motion enriches the understanding of the vortex dynamics in superlattice superconductors, which could contribute to the advent of future superconducting vortex-based devices.

## Acknowledgements

X.R.W. acknowledges support from the Academic Research Fund Tier 2 (Grant No. MOE-T2EP50210-0006 and MOE-T2EP50220-0016) and Tier 3 (Grant No. MOE2018-T3-1-002) from Singapore Ministry of Education, and Agency for Science, Technology and Research (A*STAR) under its AME IRG grant (Project No. A20E5c0094). We thank Allen Jian Yang for his assistance in sample fabrication, we also thank the fruitful discussion from W.H. Tian and Peng Song. The devices are patterned using TTT-07-Uvlitho from TuoTuo Technology, China.

**Author contributions**

X.R.W. conceived the idea and supervised the experiments. L.Z., B.S. and Q.Z. synthesized the crystal. L.Z., B.S. and Q.Z. performed the structural characterization and analysis. S.L. and A.J.Y. carried out the device fabrication. S.L. and K.H. performed the low-temperature measurements. S.L., K.H., C.Y., T.X. and L.Y. analyzed the data. S.L. and X.R.W. wrote the paper with help from all other authors.

**Competing interests**

The authors declare no competing interests.